\documentclass[12pt,a4paper]{article}
\usepackage[dvips]{graphicx}
\newcommand{\be}{\begin{eqnarray} }
\newcommand{\ee}{\end{eqnarray} }
\newcommand{\beq}{\begin{equation} }
\newcommand{\eeq}{\end{equation} }

\newcommand{\ab}{{(\alpha,\beta)}}

\begin{document}
\begin{center}
        {\large
        An approach to NLO QCD analysis of the semi-inclusive DIS data with modified Jacobi polynomial expansion method
        }
\end{center}
\begin{center}
\vskip 1.5cm
{ \large A.N.~Sissakian}
\footnote{E-mail address: sisakian@jinr.ru},  
{\large O.Yu.~Shevchenko}
\footnote{E-mail address: shevch@nusun.jinr.ru},
{\large O.N.~Ivanov}
\footnote{E-mail address: ivon@jinr.ru}\\
\vspace{1cm}
{\it Joint Institute for Nuclear Research}
\end{center}
\begin{abstract}
It is proposed 
the modification of the Jacobi polynomial expansion method (MJEM) which is
based on the application  of the truncated  moments instead of the full ones. This allows 
to reconstruct with a high precision the local quark helicity distributions even for 
the narrow accessible for measurement Bjorken $x$ region using as an input 
only four first moments extracted
from the data in NLO QCD.
It is also proposed 
the variational (extrapolation) procedure allowing to reconstruct the distributions outside 
the accessible Bjorken $x$ region using the distributions obtained with MJEM in the accessible region.
The numerical calculations encourage one that the proposed 
variational (extrapolation) procedure could be  applied 
to estimate  the full first (especially important) 
quark moments.
\end{abstract}
The extraction of the quark helicity distributions is one of the main tasks of 
the semi-inclusive deep inelastic scattering (SIDIS) experiments (HERMES \cite{hermes}, 
COMPASS \cite{compass}) with the polarized beam and target.
At the same time it was argued \cite{our1} that to obtain the reliable
distributions at relatively low average $Q^2$ available to the modern
SIDIS experiments\footnote{For example, HERMES  data \cite{hermes} on semi-inclusive asymmetries 
is obtained at $Q^2_{\rm average}=2.5GeV^2$. }, 
the leading order (LO) analysis is not sufficient  and next to leading order
analysis (NLO) is necessary. In ref. \cite{our2} the procedure allowing the direct extraction from the SIDIS data 
of the first moments of the quark helicity distributions in NLO QCD was proposed.
However, in spite of the special importance of the first moments, it is certainly very
desirable to have the procedure of reconstruction in NLO QCD of the 
polarized densities themselves. However, it is extremely difficult to  extract the local
in $x_B$ distributions directly,  because of the
double convolution product entering the NLO QCD expressions for semi-inclusive asymmetries 
(see \cite{our2} and references therein). On the other hand, operating just  as in ref. \cite{our2},
one can directly extract not only the first moments, but the Mellin moments of any required order.
The simple extension of the procedure proposed in ref. \cite{our2} gives for the n-th moments
$
\Delta_n q\equiv \int_0^1dx\, x^{n-1} q(x)
$ of the valence distributions
the equations
\be
\label{fmain}
\Delta_n u_V=\frac{1}{5}\frac{{\cal A}_p^{(n)}+{\cal A}_d^{(n)}}{L_{(n)1}-L_{(n)2}};
\quad  
\Delta_n d_V=\frac{1}{5}\frac{4
{\cal A}_d^{(n)}-{\cal A}_p^{(n)}}{L_{(n)1}-L_{(n)2}},
\ee
where all quantities in the right-hand side are the same as in ref. \cite{our2} (see Eqs. (18-23)) with the replacement
of $\int_0^1 dx$ by $\int_0^1dx\, x^{n-1}$.

It should be noticed that in reality one can measure the asymmetries only in the restricted
$x_B$ region $a<x<b$, so that the approximate equations for the truncated moments
\be
\label{qtrunc}
\Delta'_n q\equiv \int_a^bdx\, x^{n-1} q(x)
\ee
of the valence distributions have a form (\ref{fmain}) 
with the replacement of the full integrals
by the sums over bins covering accessible $x_B$ region $a<x<b$, so that
\be
\label{summ}
{\cal A}_p^{(n)}\simeq \sum_{i=1}^{N_{bins}} x^{n-1}\Delta x_i\,  A_p^{\pi^+-\pi^-}(x_i)
{\Bigl |}_Z
(4u_V-d_V)(x_i)\int_Z^1 dz_h[1+\otimes \frac{\alpha_s}{2\pi}
\tilde{C}_{qq}\otimes]
(D_1-D_2),
\ee
and analogously for ${\cal A}_d^{(n)}$.

Thus, one can directly extract from the data the n-th Mellin moments of valence distributions.
The question arises: is it sufficient to reconstruct the local in $x_B$ distributions?

There exist several methods   allowing to reconstruct the local in $x_B$ quantities
(like structure functions, polarized and unpolarized quark distributions, etc)
knowing their n-th Mellin moments. All of them use the expansion of the local
quantity in the series over the orthogonal polynomials (Bernstein,  Legendre, Jacobi, etc).
The most successful in applications (reconstruction of the local distributions
from the evolved with GLAP moments and investigation of $\Lambda_{\rm QCD}$) occurred the
Jacobi polynomial expansion method (JEM)
proposed in the pioneer work by Parisi and Sourlas \cite{parisi}
and elaborated\footnote{JEM with respect to polarized quark densities
was first applied in ref. \cite{sidorov2}.} in refs. \cite{barker} and \cite{sidorov}.
Within JEM the local in $x_B$ functions (structure functions or quark distributions) are expanded in the
double series  over the Jacobi polynomials and Mellin moments:
\be
\label{sj}
F(x)\simeq F_{N_{max}}(x)=x^\beta (1-x)^\alpha\sum_{k=0}^{N_{max}}\Theta_k^\ab(x)\sum_{j=0}^kc_j^{(k),\ab}M[j+1],
\ee
where $N_{max}$ is the number 
of moments left\footnote{Expansion (\ref{sj}) becomes exact
when $N_{max}\rightarrow\infty$. However, the advantage of JEM is that even truncated series
with small number of used  moments $N_{max}$ and properly fixed parameters $\alpha,\beta$ 
gives the good results (see, for example, \cite{sidorov}).} in the expansion.
For what follows it is of importance that the moments entering Eq. (\ref{sj}) are the {\it full} moments,
i.e., the integrals over the entire $x_B$ region $0<x<1$
$
\label{fullmoment}
M[j]=\int_0^1dx x^{j-1} F(x).
%\nonumber
$
Until now nobody investigated the question of applicability of JEM 
to the rather narrow $x_B$ region available
to the modern polarized SIDIS experiments.
So, let us try to apply JEM to the reconstruction of $\Delta u_V(x)$ and  $\Delta d_V(x)$ in
the rather narrow $x_B$ region\footnote{We choose here the most narrow HERMES  
$x_B$ region where the difference between JEM and its modification MJEM (see below) application  becomes 
especially impressive. However, even with the more wide $x_B$ region 
(for example, COMPASS  \cite{compass} region $0.003<x<0.7$) it is of importance to avoid the additional
systematical errors caused by the replacement of the full (unaccessible) moments 
in JEM (\ref{sj}) by the accessible  truncated moments.
} 
$ 
\label{region}
a=0.023<x<b=0.6
$
available to HERMES,
and to investigate
is it possible to safely replace the full moments by the truncated ones. 
To this end we perform the simple
test. We choose\footnote{Certainly, one can choose for testing any other parametrization.} GRSV2000{\bf NLO} (symmetric sea) parametrization \cite{grsv2000} at $Q^2=2.5\,GeV^2$.
Integrating the parametrization over the HERMES $x_B$ region we then calculate twelve truncated moments 
of the $u$ and $d$ valence distributions given by Eq. (\ref{qtrunc}) with $a=0.023,b=0.6$. 
Substituting these  moments in expansion
(\ref{sj}) with $N_{max}=12$, we look for optimal values of parameters $\alpha$ and $\beta$
corresponding to the minimal deviation of reconstructed curves for $\Delta u_V(x)$ and $\Delta d_V(x)$
from the input (reference) curves corresponding to input parametrization. To find these
optimal values $\alpha_{opt}$ and $\beta_{opt}$ we use the program MINUIT \cite{minuit}.
The results are presented in Fig. 1, 
where one can see that the curves strongly differ from each other      
even for the high number of used moments $N_{max}=12$.
Thus, the substitution of truncated moments instead of exact ones in the expansion
(\ref{sj}) is a rather crude approximation at least for HERMES $x_B$ region.
Fortunately it is possible to modify the standard JEM in a such way that new
series contains the truncated moments instead of the full ones. The new expansion 
looks as (see the Appendix)
\be
\label{mj}
&  & F(x)\simeq F_{N_{max}}(x)= \left(\frac{x-a}{b-a}\right)^\beta\left(1-\frac{x-a}{b-a}\right)^\alpha\nonumber\\
 & &  \times  \sum_{n=0}^{N_{max}} \Theta_n^\ab\left(\frac{x-a}{b-a}\right)
\sum_{k=0}^n c_{nk}^\ab
\frac{1}{(b-a)^{k+1}}\sum_{l=0}^k\frac{k!}{l!(k-l)!}M'[l+1](-a)^{k-l},
\ee
where we introduce the notation (c.f. Eq. (\ref{qtrunc}))
\be
\label{trm}
M'[j]\equiv M'_{[a,b]}[j]\equiv\int_a^bdx x^{j-1} F(x)
\ee
for the moments truncated to accessible for measurement $x_B$ region.
It is of great importance that now in the expansion enter not the full (unavailable) but the truncated (accessible) moments.
Thus, having at our disposal few first truncated moments extracted in NLO QCD (see Eqs. (\ref{fmain})),
and using MJEM (Eq. (\ref{mj})), one can reconstruct  the local distributions in the accessible for measurement
$x_B$ region.

To proceed let us clarify the important question about the boundary distortions.
The deviations of reconstructed with MJEM, Eq. (\ref{mj}), $F_{N_{max}}$ from F
near the boundary points are unavoidable since MJEM  is correctly defined in the entire
region $(a,b)$ except for the small vicinities of boundary points (see the Appendix). 
Fortunately,  $F_{N_{max}}$ and F are in very good agreement in the practically
entire accessible $x_B$ region,
while the boundary distortions are easy identified  and controlled since they are very sharp
and hold in very small vicinities of the boundary points (see Figs. 2-5 below). Performing the 
extrapolation outside the accessible $x_B$ region one just should cut off these
unphysical boundary distortions (see below).
%Notice that for the 
%procedure of extrapolation outside the accessible $x_B$ region one just should cut off these
%unphysical boundary distortions (see below).

Let us check how well MJEM works. To this end let us repeat the simple exercises
with reconstruction of the known GRSV2000{\bf NLO} (symmetric sea)
parametrization and compare the results of $\Delta u_V(x)$
and $\Delta d_V(x)$ reconstruction with the usual JEM and the proposed MJEM. 
To control the quality of reconstruction we introduce the parameter\footnote{
Calculating  $\nu$ we just cut off the boundary
distortions which hold for MJEM in the small vicinities of the boundary points (see the Appendix),
and decrease the integration region, respectively.
To be more precise, one can apply after cutting  some extrapolation to the boundary points.
However, the practice shows that  the results on $\nu$ calculation are practically insensitive
to the way of extrapolation since the widthes of the boundary distortion regions are very small (about $10^{-3}$).
} 
\be
\nu=\frac{\int_a^b dx|F_{reconstructed}(x)-F_{reference}(x)|}{|\int_a^b dxF_{reference}|}\cdot 100\%,
\ee
where $F_{reference}(x)$ corresponds to the input parametrization and $F_{reconstructed}(x)\equiv F_{N_{max}}(x)$ in Eq. (\ref{mj}).
We first perform the reconstruction with very high number of moments 
$N_{max}=12$ (the maximum number of moments used with standard JEM in literature -- see \cite{sidorov} 
and references therein) 
and then with small number $N_{max}=4$. Notice that the last
choice $N_{max}=4$
is especially important because of peculiarities of the data on asymmetries
provided by the SIDIS experiments.
Indeed, the number of used moments should be as small as possible
because 
first, the relative error $|\delta (M'[j])/M'[j]|$ on $M'[j]$ becomes higher with increase of $j$ and 
second, the high moments becomes very sensitive to the replacement of integration by the sum
over the bins. The results of $\Delta u_V(x)$ and $\Delta d_V(x)$ reconstruction
with  MJEM at $N_{max}=12$ and with both JEM and MJEM (in comparison) at $N_{max}=4$,
are presented by Figs. 2 and 3.
It is seen (see Fig. 2) that  for $N_{max}=12$ MJEM,  on the contrary to the usual JEM (see Fig. 1), 
gives the excellent agreement between the  reference and reconstructed curves. In the case
$N_{max}=4$ the difference in quality of reconstruction between JEM and MJEM (see Fig. 3) becomes 
especially impressive. 
While for standard JEM the reconstructed and reference curves strongly 
differ from each other, the respective curves for MJEM are in a good 
agreement. Thus, one can conclude that dealing with the truncated, available to measurement,
$x_B$ region one should apply the proposed modified JEM to obtain
the reliable results on the local distributions.

Until now we looked for  the optimal values of parameters $\alpha$ and $\beta$ entering 
MJEM using explicit form of the reference curve (input parametrization). Certainly, in reality
we have no any reference curve to be used for optimization. However, one can extract from the data
in NLO QCD the first few moments (see  Eqs. (\ref{fmain})). Thus, we need some
criterium  of MJEM optimization which would use for optimization of $\alpha$ and $\beta$
only the known (extracted) moments entering MJEM.

On the first sight it seems to be natural to find the optimal values of $\alpha$ and $\beta$ minimizing
the difference of {\it reconstructed} with MJEM and input\footnote{In practice one should
reconstruct these input moments from the data using Eqs. (\ref{fmain}). The reference ``twice-truncated''
moments (\ref{trm2}) should be reconstructed from the data in the same way.} ({\it entering}
MJEM expansion (\ref{mj}))  moments.
However, it is easy to prove\footnote{It can be proved by analogy with the case of the usual JEM,
where Eq. (\ref{meq}) with $[a,b]=[0,1]$ holds (see, for example, \cite{sidorov}).} 
that this difference is equal to zero identically: 
\be
\label{meq}
M'_{[a,b]}[n]{\Biggl |}_{\rm reconstructed}\equiv M'_{[a,b]}[n]{\Biggl |}_{input},\quad n\le N_{max},
\ee
i.e. all reconstructed moments with $n\le N_{max}$ are identically equal to the respective
input moments {\it for any} $\alpha$ and $\beta$. Fortunately, we can use for comparison
the reference  ``twice-truncated'' moments
\be
\label{trm2}
M''[n]\equiv M''_{[a+a',b-b']}[n]\equiv\int_{a+a'}^{b-b'} dx\, x^{n-1} F(x)\quad
(a<a+a'<b-b'<b),
\ee
i.e. the integrals over the region less than the integration region $[a,b]$ for the  
``once-truncated''  moments $M'_{[a,b]}$ entering MJEM (\ref{mj}). The respective optimization criterium 
can be written in the form
\be
\label{criterium}
\sum_{j=0}^{N_{max}}{\Bigl |}M''_{\rm (reconstructed)}[j]-M''_{\rm (reference)}[j]{\Bigl |}=min.
\ee
The ``twice truncated'' reference moments should be extracted in NLO QCD from the data in the same way
as the input (entering MJEM (\ref{mj})) ``once truncated'' moments. In reality
one can obtain ``twice-truncated'' moments using Eqs. (\ref{fmain}) and removing, for example, first and/or last bin from the
sum in Eq. (\ref{summ}).

Let us now check how well the optimization criterium (\ref{criterium})
works. 
To this end we again perform the simple numerical
test. We choose GRSV2000{\bf NLO} parametrization at $Q^2=2.5\,GeV^2$  with both broken and  symmetric sea
scenarios. We then calculate four first ``once-truncated'' and four first ``twice-truncated'' moments 
defined by Eqs. (\ref{trm}) and (\ref{trm2}), and substitute them in the optimization criterium 
(\ref{criterium}). To find the optimal values of $\alpha$ and $\beta$ we use the MINUIT \cite{minuit}
program. The results are presented by Fig. 4.
It is seen that the optimization criterium works well for both symmetric and broken
sea scenarios.

Thus, one can conclude that MJEM can be successfully applied for reconstruction
of the local distributions knowing only  few first truncated Mellin moments. 
Notice, however, that by construction MJEM reproduces the local distributions only in the
accessible for measurement $x_B$ region. The question arises: could one 
attempt to reconstruct the local distributions outside the accessible region (i.e. to perform
extrapolation) using
the obtained with MJEM distributions %within the available to measurement region as
as an input? To this end we propose to solve the following variational task.
We apply MJEM, Eq. (\ref{mj}), to the maximally\footnote{
For a moment, we restrict ourselves by the $x_B$ region $[a_{min}=10^{-4},b_{max}=1]$
which is typical for the most  known parametrizations on the quark helicity distributions.
}  extended $x_B$ region $[a_{min},b_{max}]$
replacing the moments $M'_{[a_{min},b_{max}]}[j]$ by  
$M'_{[a,b]}[j]+\epsilon_j
$,
where $\epsilon_j (j=1\ldots4)$ are the free variational parameters ($\epsilon_j$ should  be considered as 
unknown ``tails'' of the full moments).
Then, using MINUIT program \cite{minuit}, one finds the parameters  $\epsilon_j$
requiring the minimal deviation of the reconstructed with $\epsilon_j$ curve
from the input (reconstructed with criterium (\ref{criterium})) curve 
inside the accessible for measurement region $[a,b]$. The reconstructed 
in this way quantities $M'_{[a,b]}[j]+\epsilon_j$ should be compared with the reference
(obtained by direct integration of the input parametrization) moments 
$M'_{[a_{min},b_{max}]}[j]{\Bigl |}_{\rm reference}$. In ideal case (ideal reconstruction
of ``tails'' $\epsilon_j$) these quantities would coincide.

Let us test this variational (extrapolation) procedure by the simple numerical exercise. 
We choose GRSV2000 {\bf NLO} parametrization (for both broken and symmetric sea scenarios) at $Q^2=2.5\,GeV^2$
as the reference one. Since the allowed \cite{grsv2000} $x_B$ region for this  parametrization
is $[10^{-4},1]$ we choose $[a_{min},b_{max}]=[10^{-4},1]$ and 
for the truncated region $[a,b]$ we again choose the accessible for HERMES $x_B$ region $[a,b]=[0.023,0.6]$. 
Notice that performing the variational (extrapolation) procedure
we cut off the boundary distortions of the curve (which enters the variational procedure as an
input) obtained with MJEM and criterium
(\ref{criterium}) inside the accessible $x_B$ region.
The results of the  variational (extrapolation) procedure application are presented by Fig 5 and Table 1.

\begin{table}
        \caption{Results of first four moments of  $\Delta u_V$ and $\Delta d_V$ reconstruction
        in the region $[a_{min}=10^{-4},b_{max}=1]$ for the GRSV2000{\bf NLO} parametrization
        for both symmetric (top) and broken sea (bottom) scenarios.}
\begin{tabular}{cccc|ccc}
        \hline\hline
         & \multicolumn{3}{c|}{$\Delta u_V(x)$} & \multicolumn{3}{c}{$\Delta d_V(x)$} \\ 
        n & $M_{[0.023,0.6]}^{\rm 'input}$ & $M_{[10^{-4},1]}^{\rm' output}$  &  $M_{[10^{-4},1]}^{\rm' reference}$& $M_{[0.023,0.6]}^{\rm 'input}$ & $M_{[10^{-4},1]}^{\rm' output}$  &  $M_{[10^{-4},1]}^{\rm' reference}$  \\ 
        \hline
1 & 0.749  &0.904  & 0.917& -0.275  & -0.362 &  -0.340\\
2 & 0.153  &0.164  & 0.167& -0.049  & -0.051 &  -0.051\\
3 & 0.047  &0.053  & 0.055& -0.013  & -0.014 &  -0.014\\
4 & 0.017  &0.021  & 0.023& -0.004  & -0.005 &  -0.005\\
        \hline\hline
1 & 0.570 & 0.609 & 0.605& -0.114 & -0.074 & -0.029\\
2 & 0.137 & 0.150 & 0.149& -0.036 & -0.038 & -0.037\\
3 & 0.044 & 0.052 & 0.052& -0.012 & -0.013 & -0.013\\
4 & 0.017 & 0.023 & 0.022& -0.004 & -0.006 & -0.005\\
\hline

\end{tabular}
\end{table}
Comparing the reconstructed and input curves (see Fig. 5) one can see that they are in a
quite good agreement while  the slight deviation should be corrected in the future by improving the
variational (extrapolation) procedure. 
Fortunately, all four moments  occur almost insensitive to this deviation. Indeed, from the Table 1
it is seen that the reconstructed full moments for both $u$ and $d$ valence  quarks and for both 
scenarios are in a good agreement with the respective reference values.
The most interesting is  that the reconstructed first moments which are of the most importance
for understanding the proton spin puzzle are in  good agreement with their reference values.

Thus, all numerical tests confirm that the proposed modification of the 
Jacobi polynomial expansion method, MJEM, allows to reconstruct with a high precision
the quark helicity distributions in the accessible for measurement $x_B$ region.
Besides, the numerical calculations demonstrate that the
application of MJEM together with the special variational (extrapolation) procedure
can allow to estimate the full first (especially important) quark moments
knowing the distributions reconstructed with MJEM in the accessible for measurement
$x_B$ region.

  The authors are grateful to R.~Bertini,
 O.~Denisov, 
 A.~Korzenev, V.~Krivokhizhin, E.~Kuraev,
 A.~Maggiora,  A.~Nagaytsev, A.~Olshevsky, 
G.~Piragino, G.~Pontecorvo,  I.~Savin, A.~Sidorov and  O.~Teryaev,
 for fruitful discussions. Two of us (O.S, O.I.) thanks RFBR grant 05-02-17748.

\begin{center}
Appendix
\end{center}
\renewcommand{\theequation}{A.\arabic{equation}}
\setcounter{equation}{0} 

The  JEM is the expansion of the $x$-dependent function (structure function or quark density)
in the series over Jacobi polynomials 
$\Theta^\ab_n(x)$ orthogonal with weight $\omega^{(\alpha,\beta)}(x)=x^\beta(1-x)^{\alpha}$ (see \cite{parisi}
-\cite{sidorov} for details):
\be
\label{a1}
F(x)=\omega^\ab\sum_{k=0}^\infty\Theta_k^\ab(x)\sum_{j=0}^kc_{kj}^{\ab}M(j+1),
\ee
where 
\be
M[j]=\int_0^1dx\, x^{j-1} F(x)
\ee
and 
\be
\label{ortho}
\int_0^1dx\omega^\ab(x)\Theta^\ab_n(x)\Theta^\ab_m(x)=\delta_{nm}.
\ee
The details on the Jacobi polynomials 
\be
\label{razlojenie}
\Theta_k^\ab(x)=\sum_{j=0}^kc_{kj}^\ab x^j
\ee
 can be found in refs. \cite{parisi} and
\cite{barker}. In practice one truncates  the series (\ref{razlojenie})
living in the expansion only finite number of moments $N_{max}$ -- see Eq. (\ref{mj}).
The experience shows \cite{sidorov} that JEM produces good results even with 
small number $N_{max}$.

The idea of modified expansion is to reexpand $F(x)$ in the series over the truncated
moments $M'_{[ab]}[j]$ given by Eq. (\ref{trm}), 
performing the rescaling $x\rightarrow a+(b-a)x$ which compress the entire region
$[0,1]$ to the truncated region $[a,b]$. To this end let us apply the following ansatz\footnote{
Notice that ansatz (\ref{e1}) (as well as the expansion Eq. (\ref{mj}) itself) is correctly defined inside the entire region $(a,b)$ except for
the small vicinities of boundary points (absolutely the same situation holds for the usual JEM, Eq. (\ref{a1}), applied
to the quark distributions in the region $(0,1)$). In practice, the respective boundary distortions are just cut off when one performs
the extrapolation procedure.
}
\be
\label{e1}
F(x)=\left(\frac{x-a}{b-a}\right)^\beta\left(1-\frac{x-a}{b-a}\right)^\alpha\sum_{n=0}^\infty\tilde f_n \Theta_n^\ab\left(\frac{x-a}{b-a}\right)
\ee
and try to find the coefficients $\tilde f_n$. Multiplying both parts of Eq. (\ref{e1})
by $\Theta_k^\ab((x-a)/(b-a))$, integrating over x in the  limits $[a,b]$ and  performing the
replacement $t=(x-a)/(b-a)$, one gets
\be
\int_a^bdx\, F(x)\Theta_k^\ab\left(\frac{x-a}{b-a}\right)=
(b-a)\sum_{n=0}^\infty{\tilde f_n}\int_0^1dt\, t^\beta(1-t)^\alpha \Theta_n^\ab(t)\Theta_k^\ab(t),
%=\frac{\tilde f_k}{b-a}
\ee
so that with the orthogonality condition Eq. (\ref{ortho})
one obtains 
\be
\label{e2}
\tilde f_n=(b-a)^{-1}\int_a^b dx\, F(x) \Theta_n^\ab\left(\frac{x-a}{b-a}\right).
\ee
Substituting Eq. (\ref{e2}) in the expansion (\ref{e1}), and using Eq. (\ref{razlojenie}) one eventually gets 
\be
\label{apfinal}
 &  & F(x)  =  \left(\frac{x-a}{b-a}\right)^\beta\left(1-\frac{x-a}{b-a}\right)^\alpha\nonumber\\
 & &  \times  \sum_{n=0}^{\infty} \Theta_n^\ab\left(\frac{x-a}{b-a}\right)
 \sum_{k=0}^n c_{nk}^\ab
 \frac{1}{(b-a)^{k+1}}\sum_{l=0}^k\frac{k!}{l!(k-l)!}M'_{[a,b]}[l+1](-a)^{k-l},
\ee
where $M'_{[a,b]}[j]$ is given by Eq. (\ref{trm}).
Truncating in the exact Eq. (\ref{apfinal}) the infinite sum over $n$ to the sum $\sum_{n=0}^{N_{max}}$
one gets the approximate equation (\ref{mj}).

\begin{figure}[htb!]
        \label{f1}
        \caption{The results of reconstruction of $\Delta u_V(x)$ ($\alpha_{opt}=8.18922$, $\beta_{opt}=-0.99$)
        and $\Delta d_V(x)$ ($\alpha_{opt}=-0.99$, $\beta_{opt}=-0.387196$)
         with the usual JEM. Solid line corresponds to input (reference) parametrization.
         Dotted line corresponds to the distributions reconstructed with JEM .}
\begin{center}
{\includegraphics[height=4cm,width=6cm]{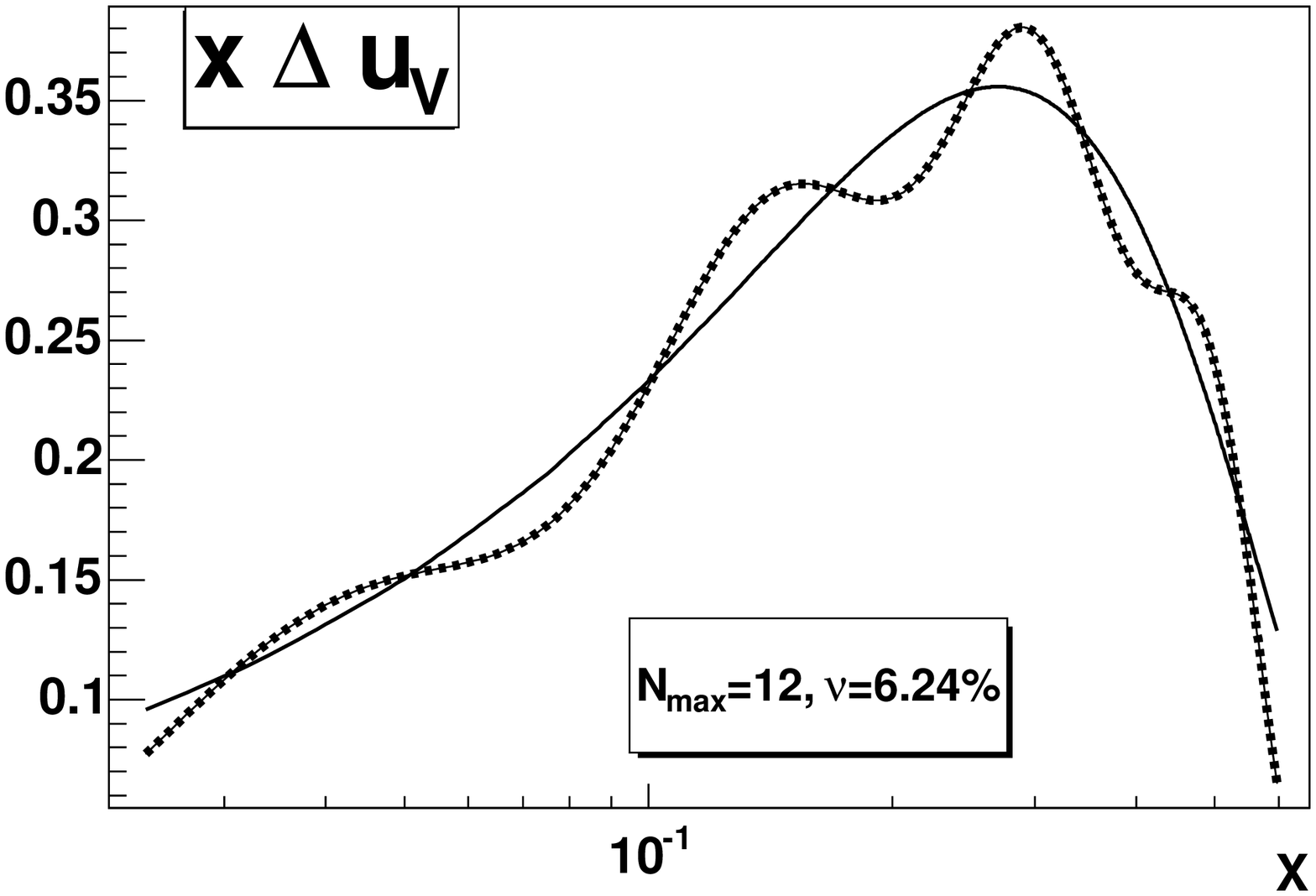}}
{\includegraphics[height=4cm,width=6cm]{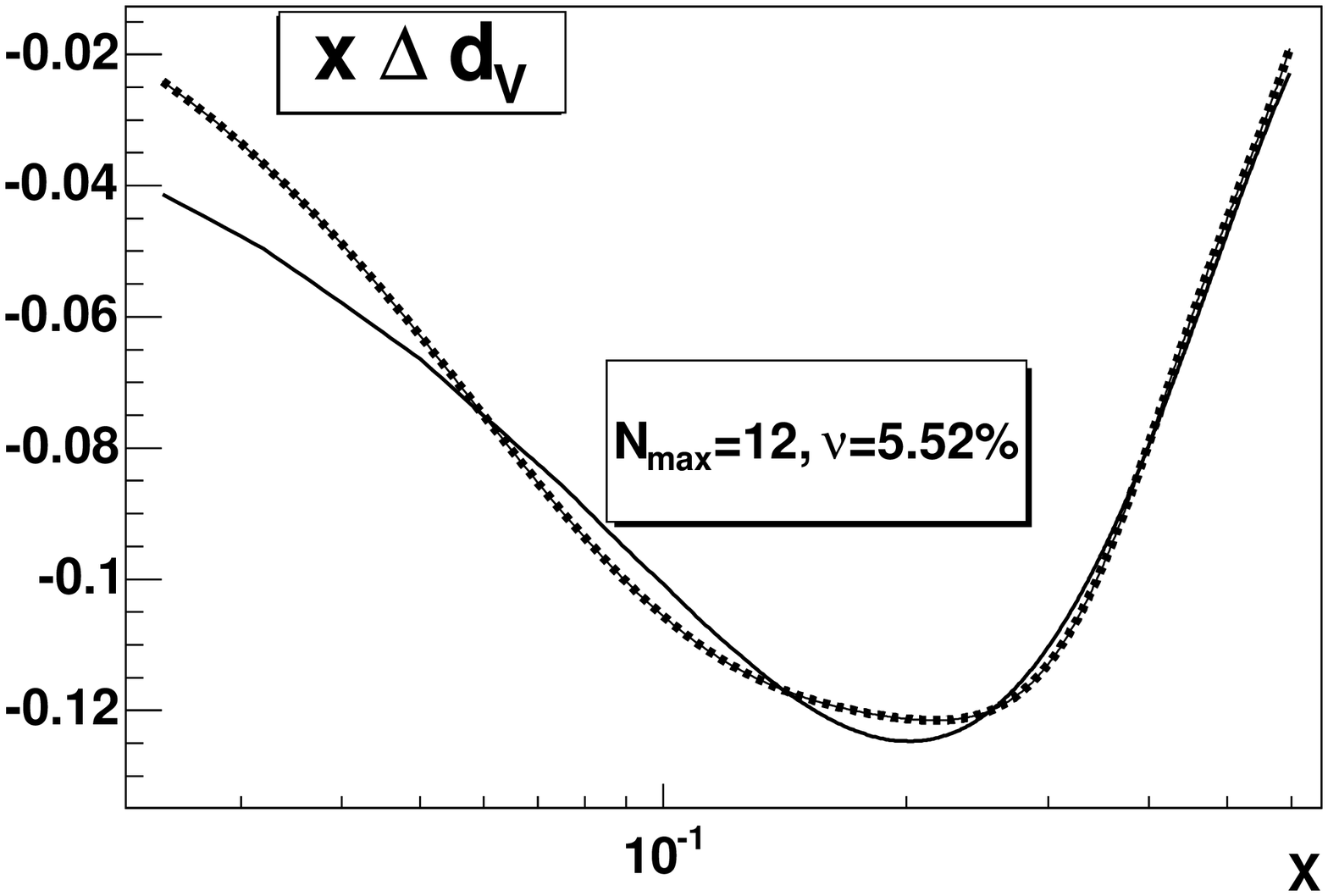}}
\end{center}
\end{figure}

\begin{figure}[htb!]
        \caption{The results of $\Delta u_V(x)$ ($\alpha_{opt}=-0.827885$, $\beta_{opt}=-0.011505$)
        and $\Delta d_V(x)$ ($\alpha_{opt}=-0.989752$, $\beta_{opt}=-0.012393$) reconstruction 
         with MJEM. Solid line corresponds to input (reference) parametrization.
         Dotted line corresponds to the distributions reconstructed with MJEM .}
\begin{center}
%{\includegraphics[height=4cm,width=6cm]{j-uv-id1-n12.eps}}
%{\includegraphics[height=4cm,width=6cm]{j-dv-id1-n12.eps}}
{\includegraphics[height=4cm,width=6cm]{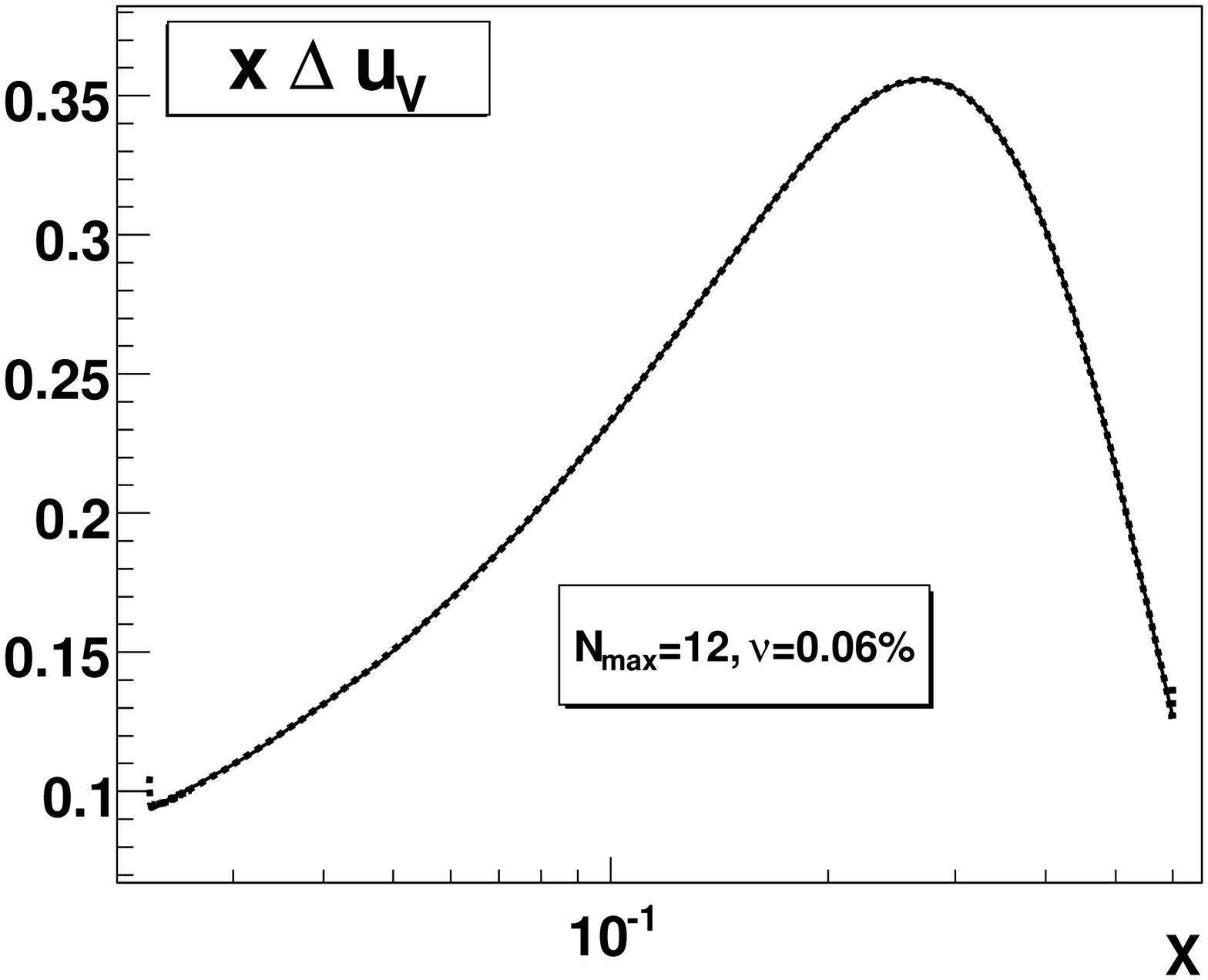}}
{\includegraphics[height=4cm,width=6cm]{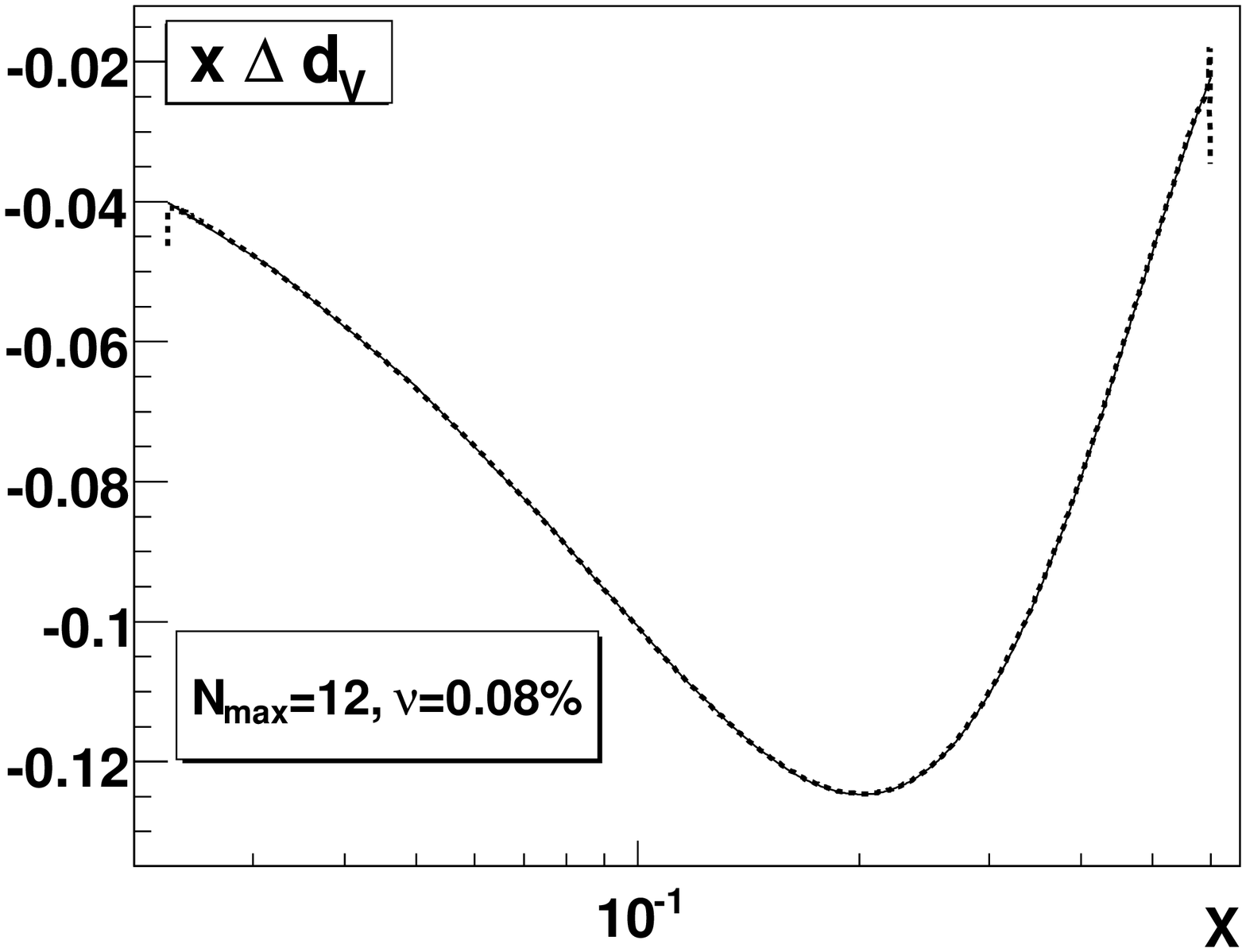}}
\end{center}
\end{figure}
\begin{figure}[htb!]
        \caption{ The top part corresponds to  
        $\Delta u_V(x)$ ($\alpha_{opt}=-0.99$, $\beta_{opt}=0.054010$)
        and $\Delta d_V(x)$ ($\alpha_{opt}=0.174096$, $\beta_{opt}=0.162567$) 
        reconstructed with the usual JEM.
        The bottom part corresponds to 
        $\Delta u_V(x)$ ($\alpha_{opt}=-0.0025869$, $\beta_{opt}=-0.071591$)
        and $\Delta d_V(x)$ ($\alpha_{opt}=0.110331$, $\beta_{opt}=-0.049255$) reconstructed with MJEM.
        Solid lines correspond to input (reference) parametrization.
        Dotted lines correspond to the distributions reconstructed with JEM (top) and MJEM (bottom).}
\begin{center}
{\includegraphics[height=4cm,width=6cm]{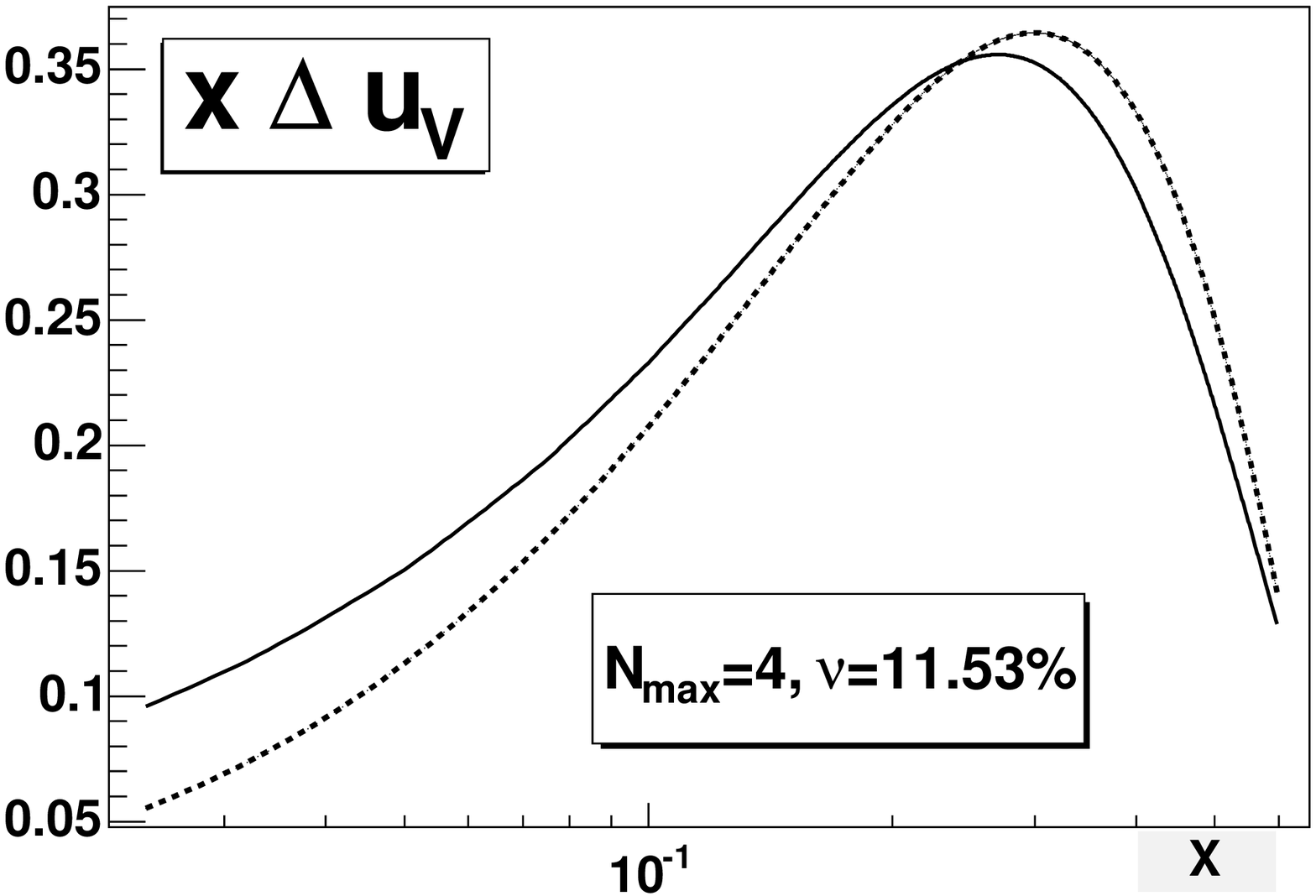}}
{\includegraphics[height=4cm,width=6cm]{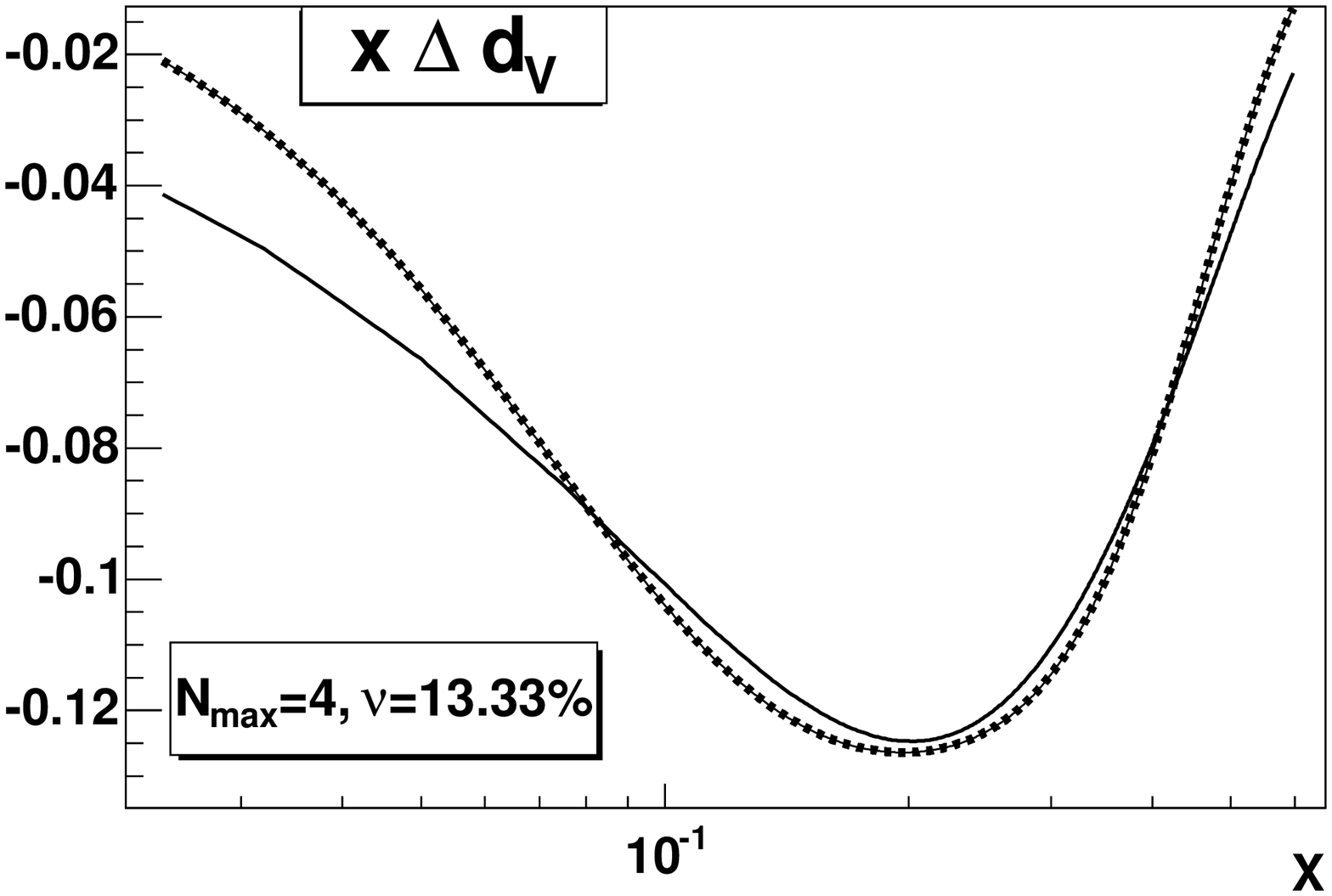}}
{\includegraphics[height=4cm,width=6cm]{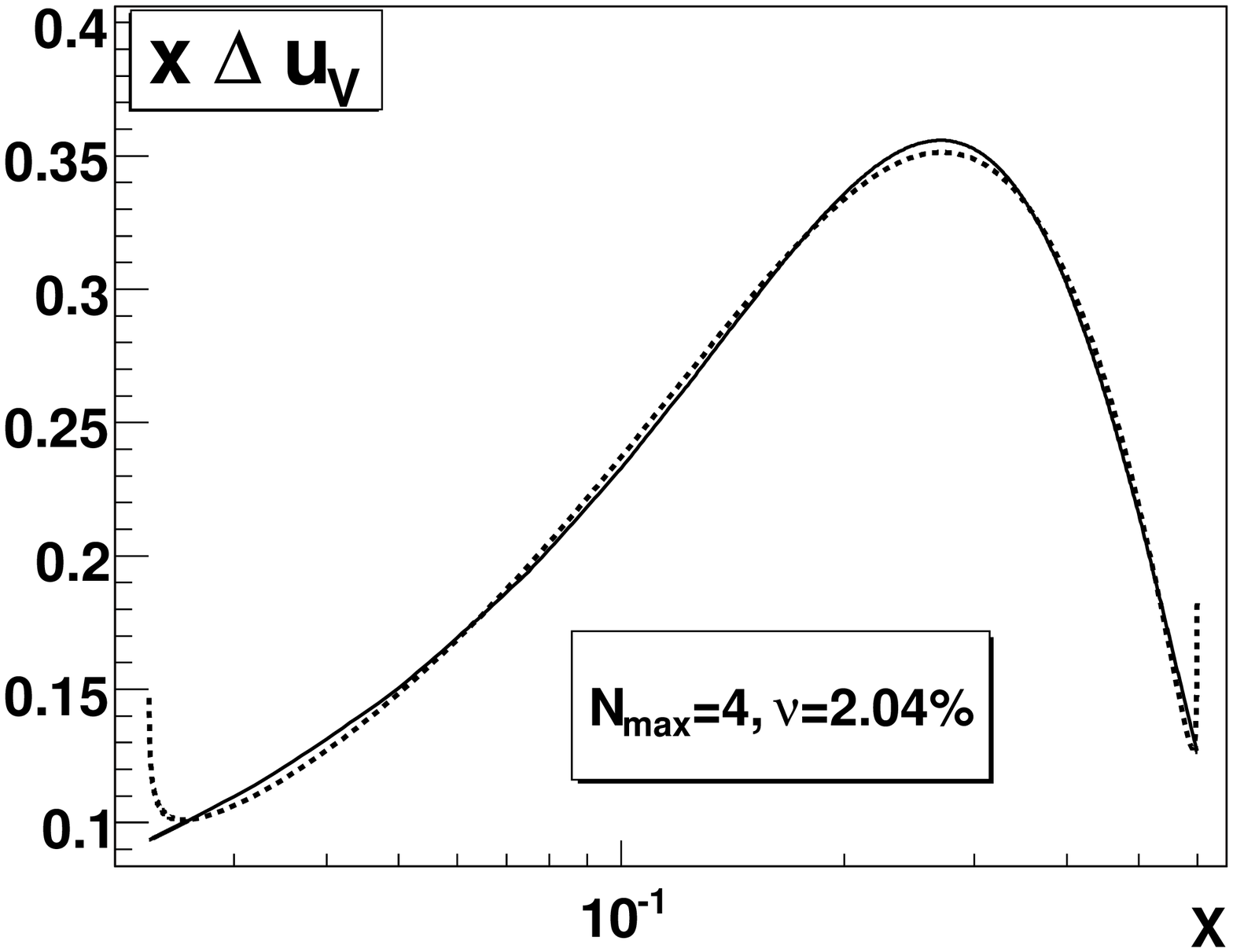}}
{\includegraphics[height=4cm,width=6cm]{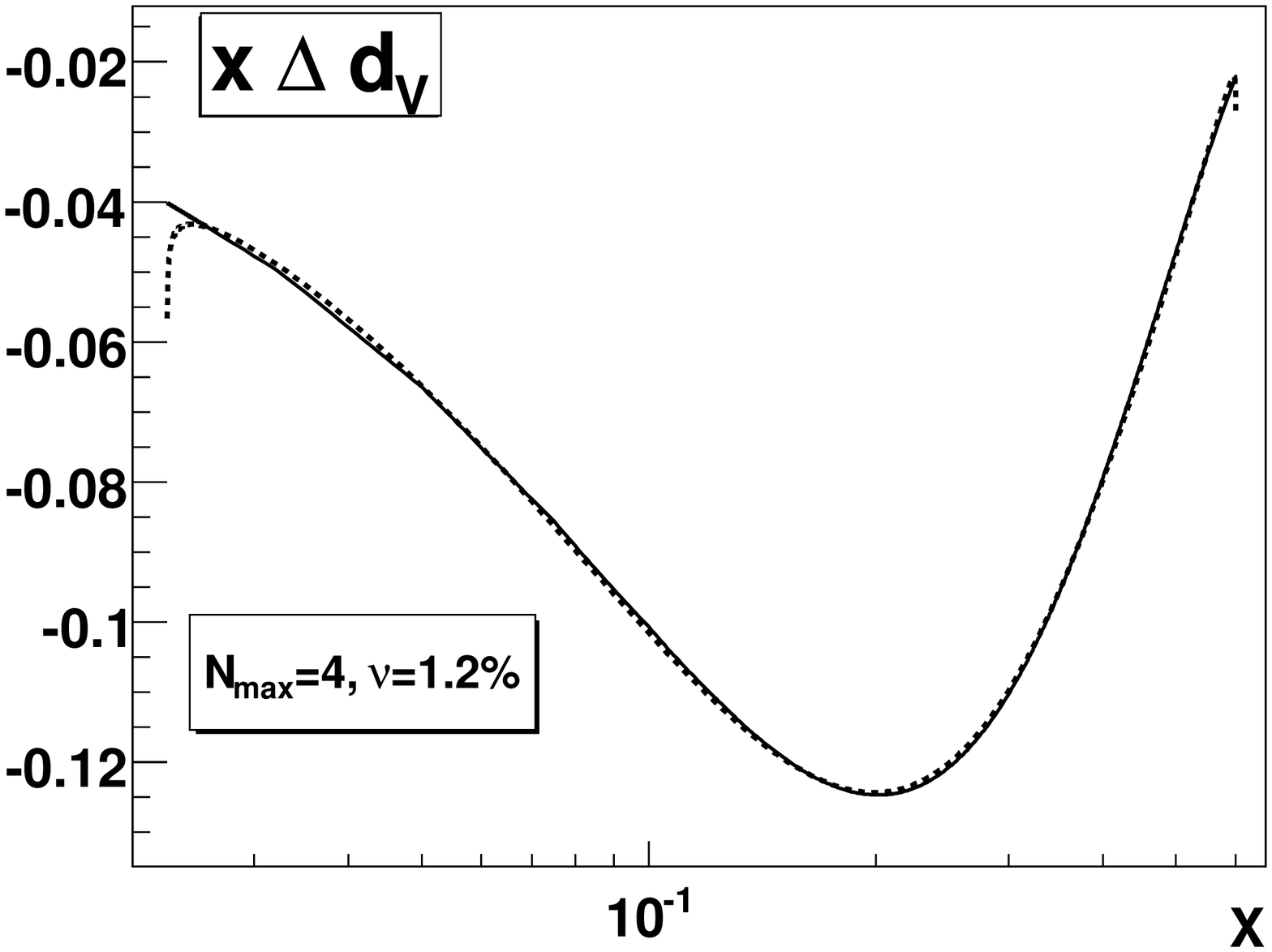}}
\end{center}
\end{figure}

\begin{figure}[htb!]
        \begin{center}
        \caption
        { 
        The results of $\Delta u_V$ and $\Delta d_V$ reconstruction   for  GRSV2000{\bf NLO} 
        parametrization for both symmetric (top) and
        broken sea (bottom) scenarios.
        Solid line corresponds to the reference curve (input parametrization).
        Dotted line is 
         reconstructed with MJEM and criterium (\ref{criterium}) 
        inside the accessible for measurement region ([0.023,0.6] here).
        Optimal values of parameters for symmetric sea scenario for $\Delta u_V$ are $\alpha_{opt}=-0.15555$,
        $\beta_{opt}=-0.097951$ and for $\Delta d_V$ are $\alpha_{opt}=-0.002750$, $\beta_{opt}=-0.07190$.
        Optimal values of parameters for broken sea scenario for $\Delta u_V$ are $\alpha_{opt}=-0.209346$,
        $\beta_{opt}=0.153417$ and for $\Delta d_V$ are $\alpha_{opt}=0.702699$, $\beta_{opt}=-0.293231$.
        }
{\includegraphics[height=4cm,width=6cm]{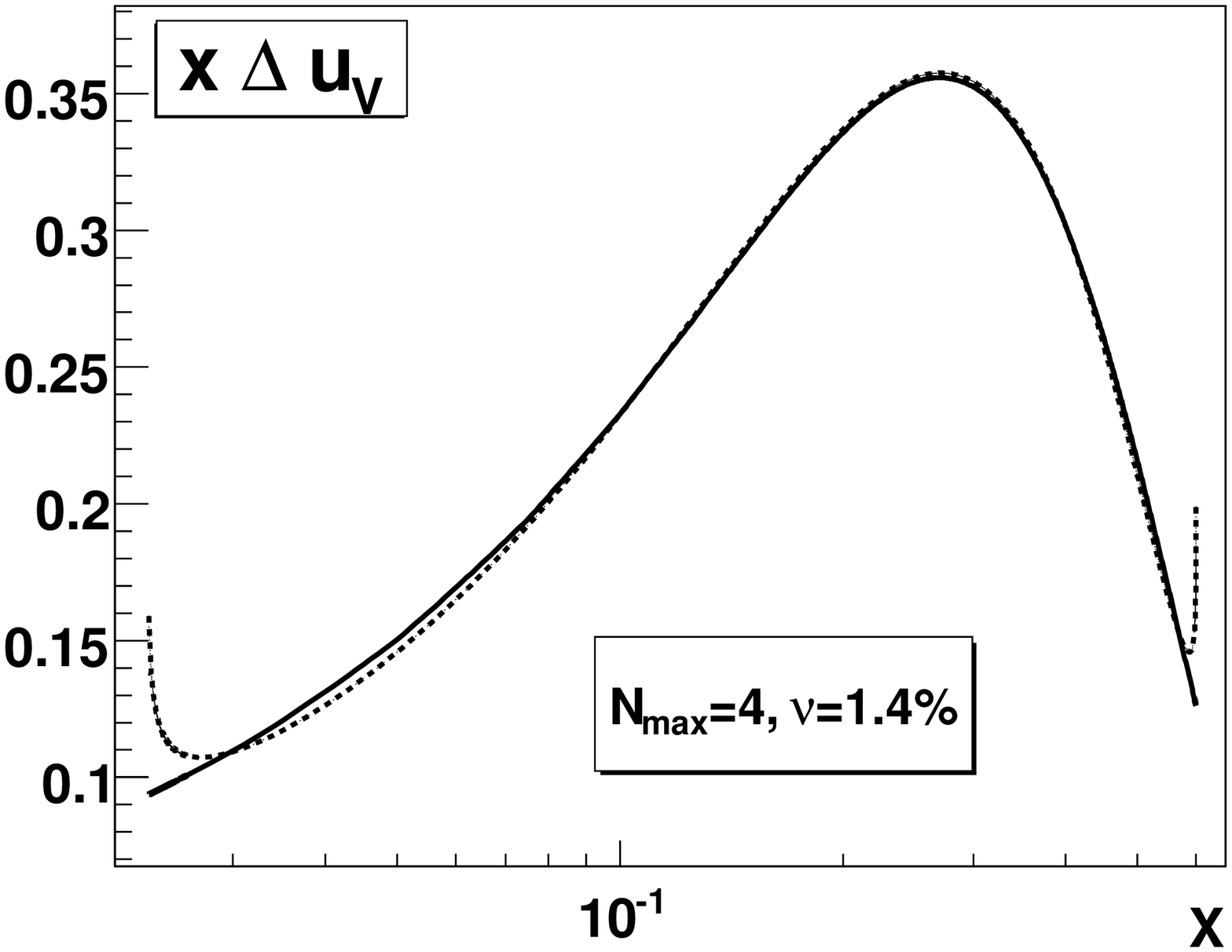}}
{\includegraphics[height=4cm,width=6cm]{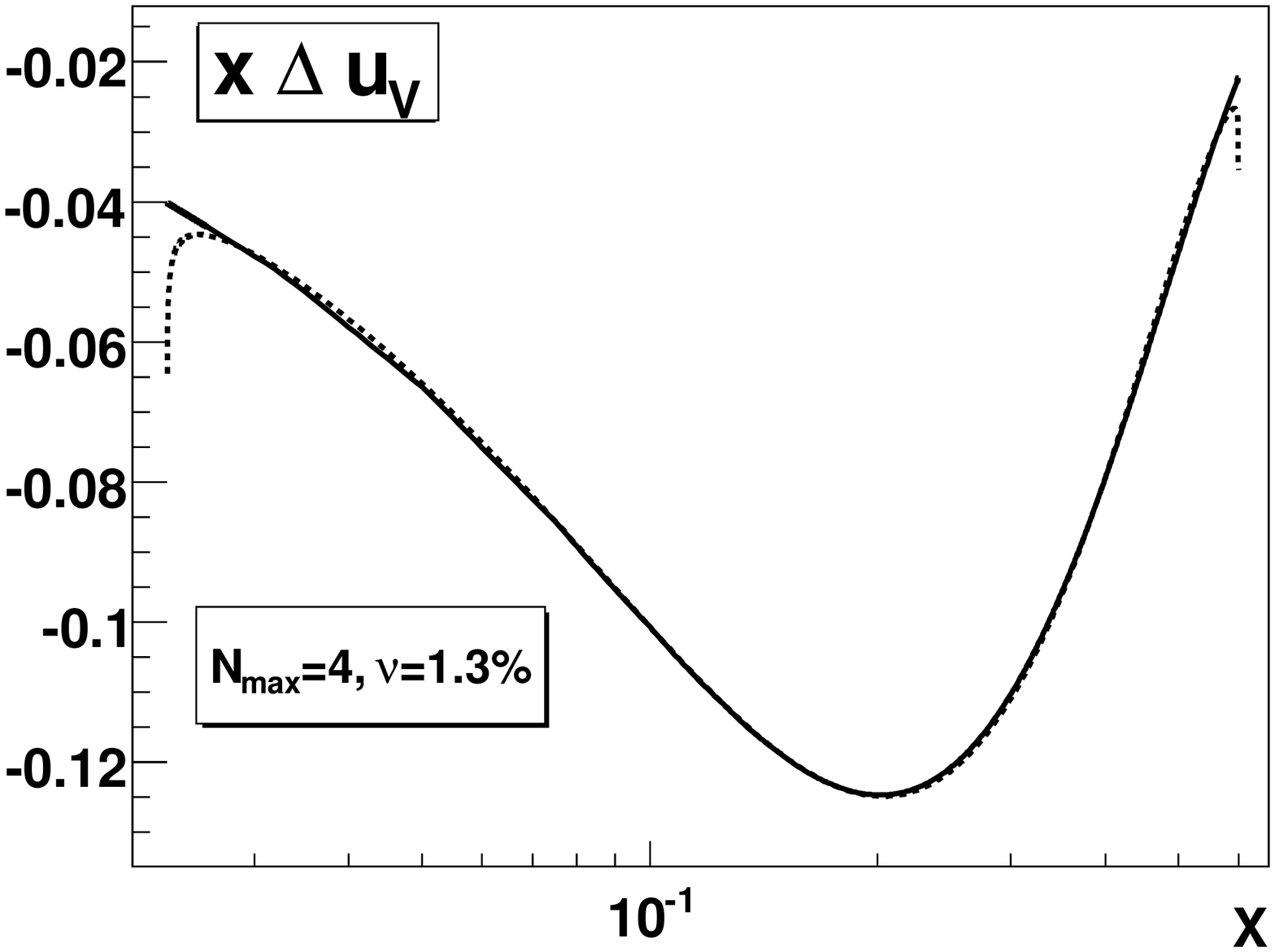}}
{\includegraphics[height=4cm,width=6cm]{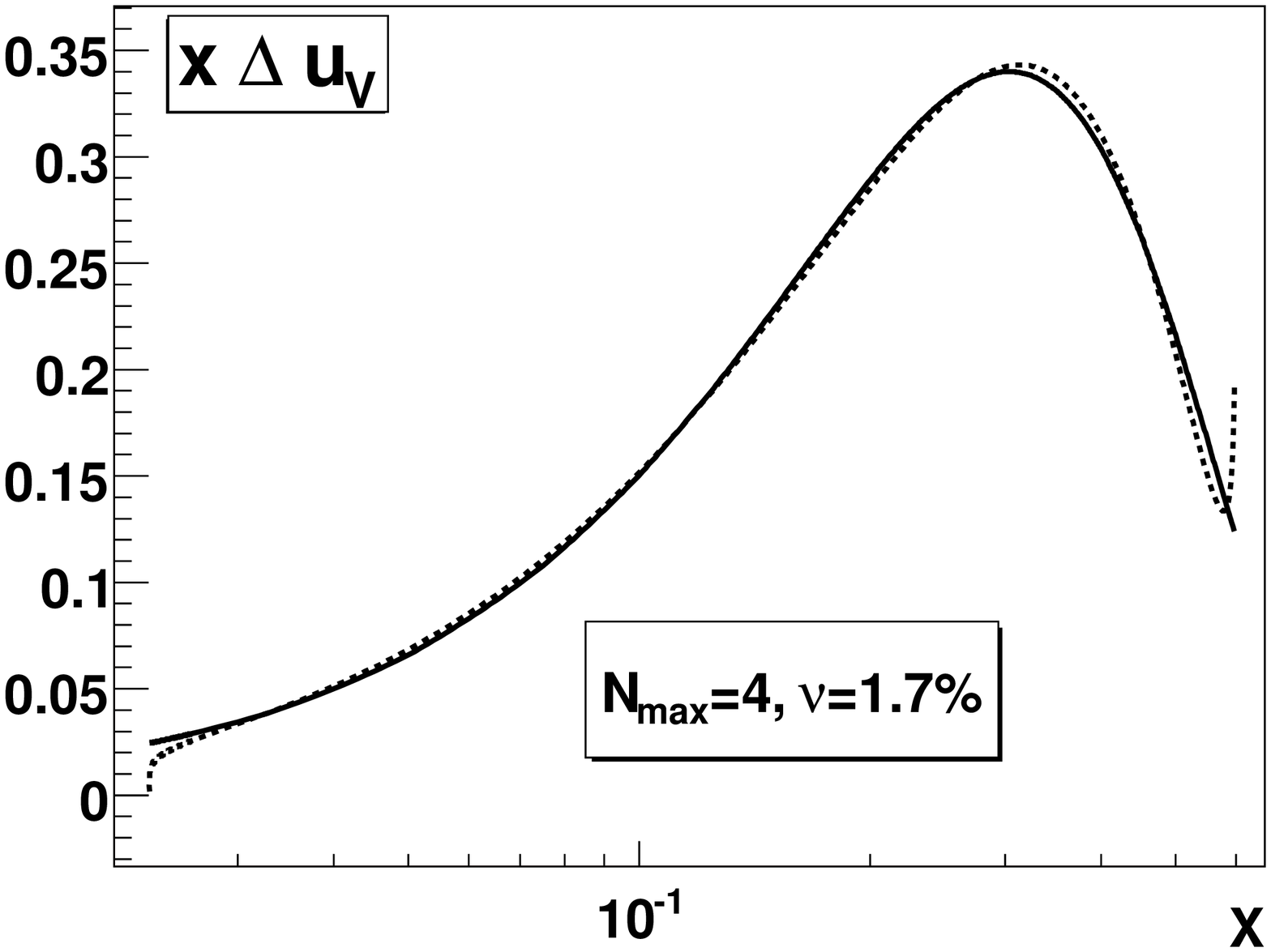}}
{\includegraphics[height=4cm,width=6cm]{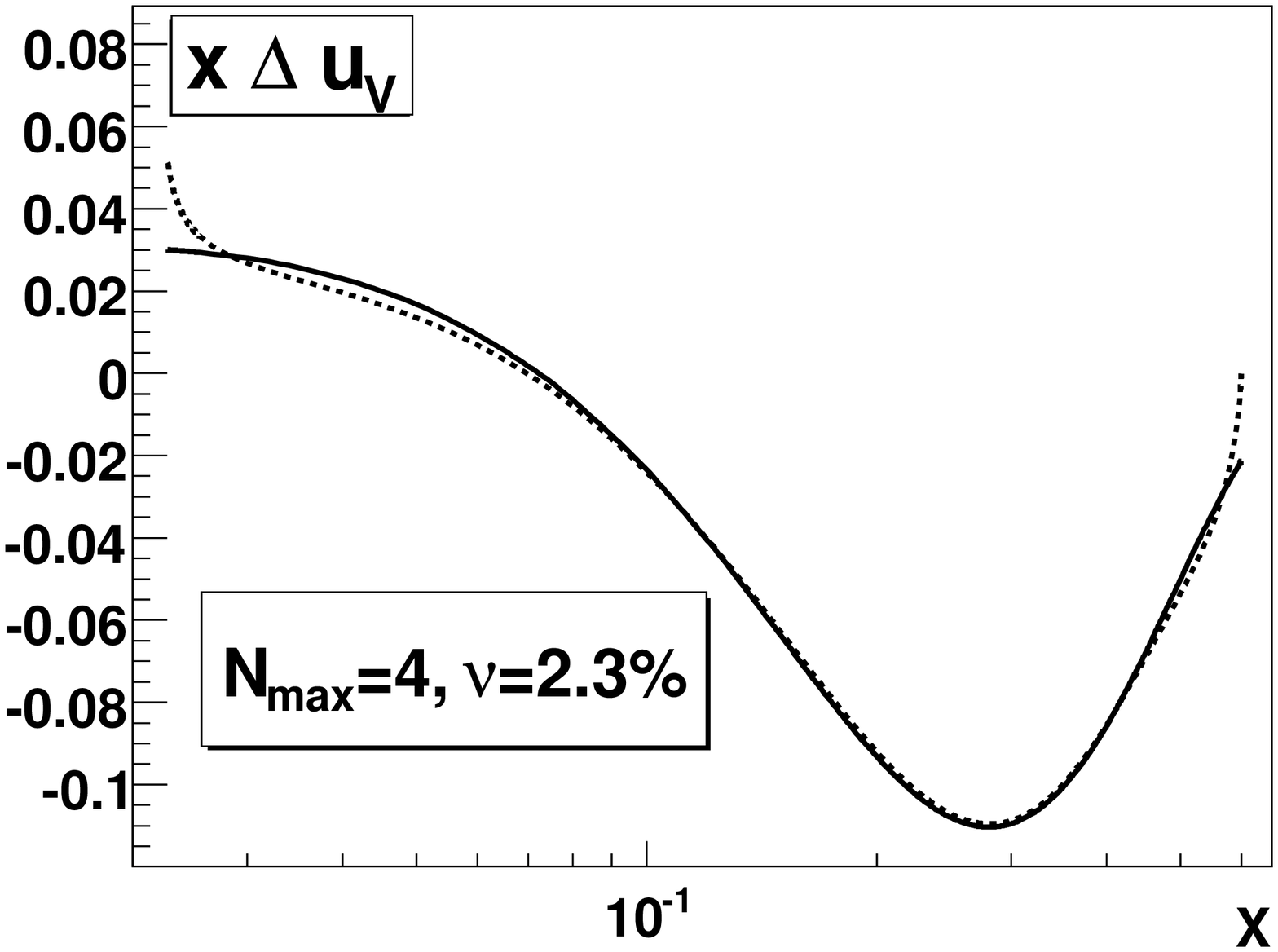}}
\end{center}
\end{figure}
\begin{figure}[htb!]
        \begin{center}
        \caption
        { 
        The results of $\Delta u_V$ and $\Delta d_V$ reconstruction
        in the region $[a_{min}=10^{-4},b_{max}=1]$
        for GRSV2000{\bf NLO} parametrization for both symmetric (top) and
        broken sea (bottom) scenarios.
        Solid line corresponds to the reference curve (input parametrization).
        Dotted line 
        corresponds to the curve reconstructed in the entire $[a_{\rm max}=10^{-4},b_{\rm max}=1]$ region
        with requirement of minimal deviation from the
        curve (bold solid line ) reconstructed with MJEM and criterium (\ref{criterium}) 
        inside the accessible for measurement region ([0.023,0.6] here).
        }
{\includegraphics[height=4cm,width=6cm]{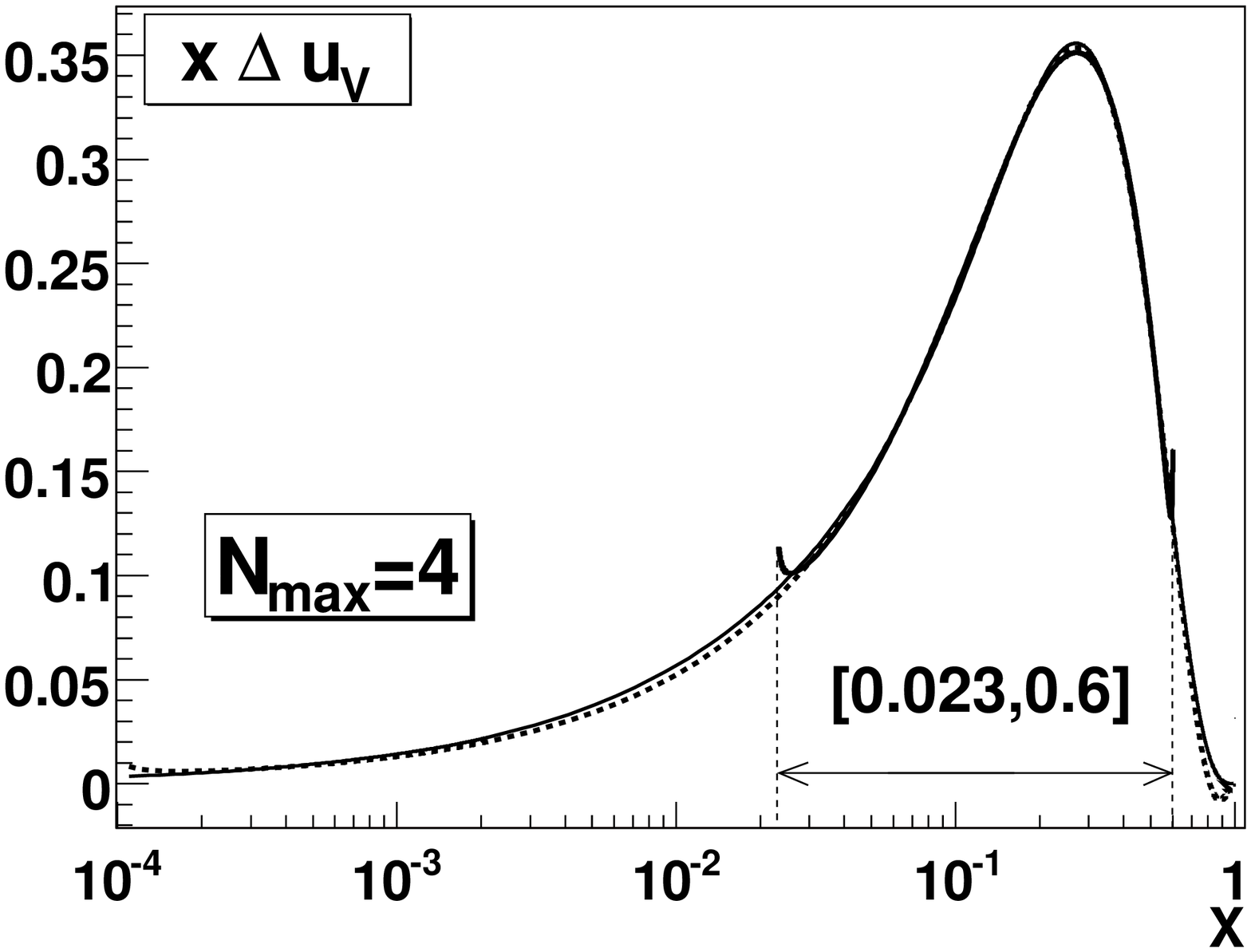}}
{\includegraphics[height=4cm,width=6cm]{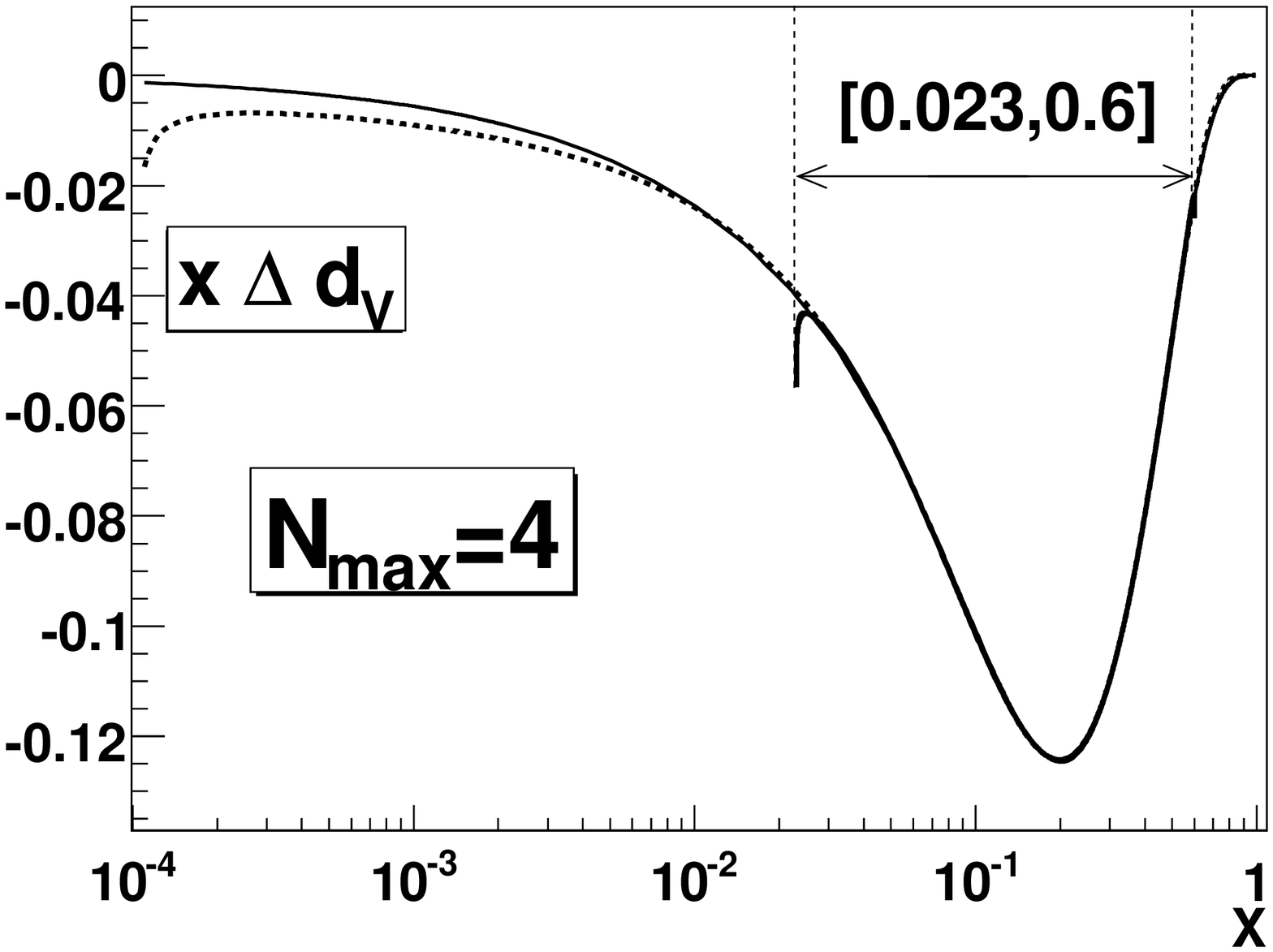}}
{\includegraphics[height=4cm,width=6cm]{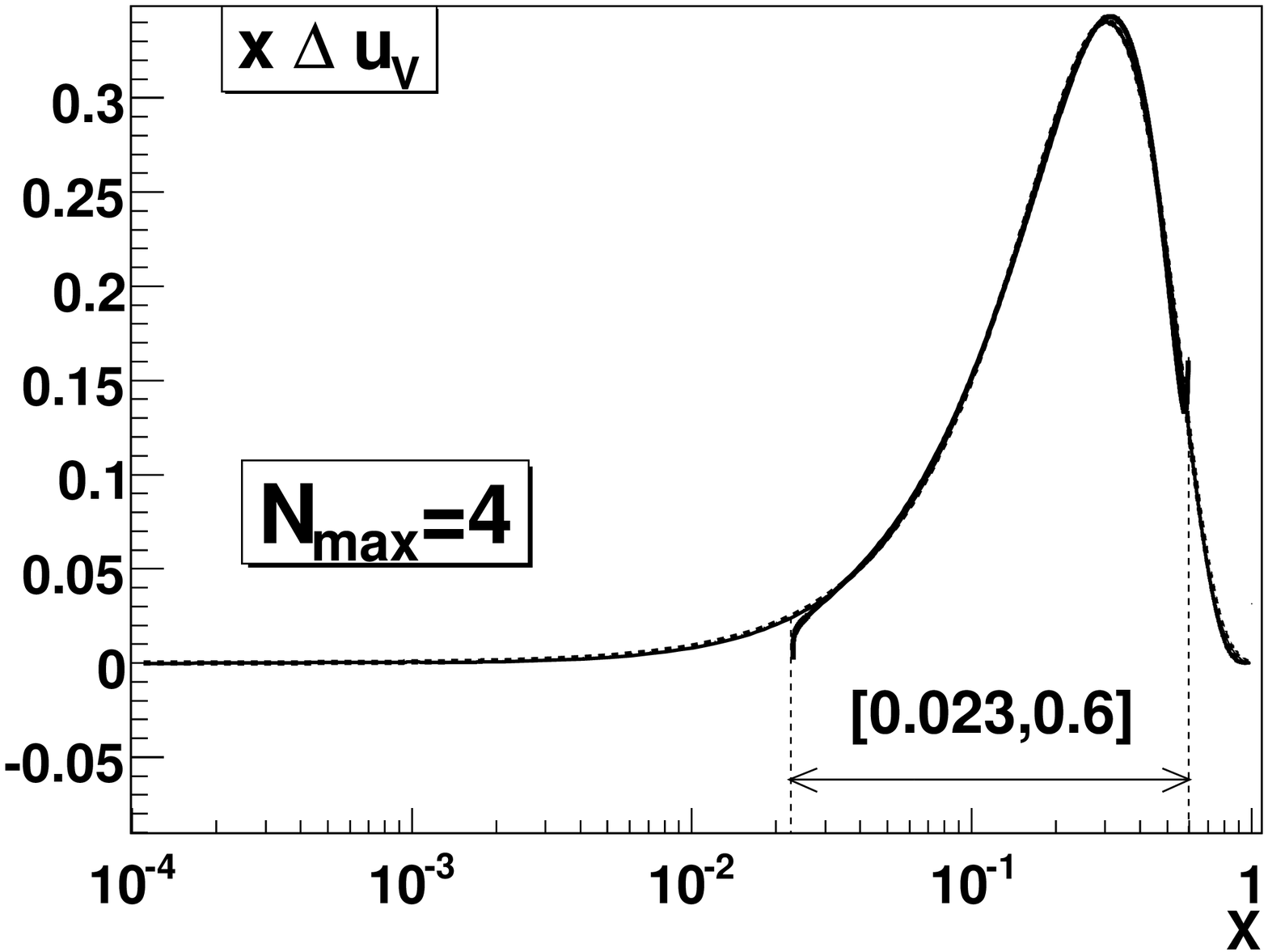}}
{\includegraphics[height=4cm,width=6cm]{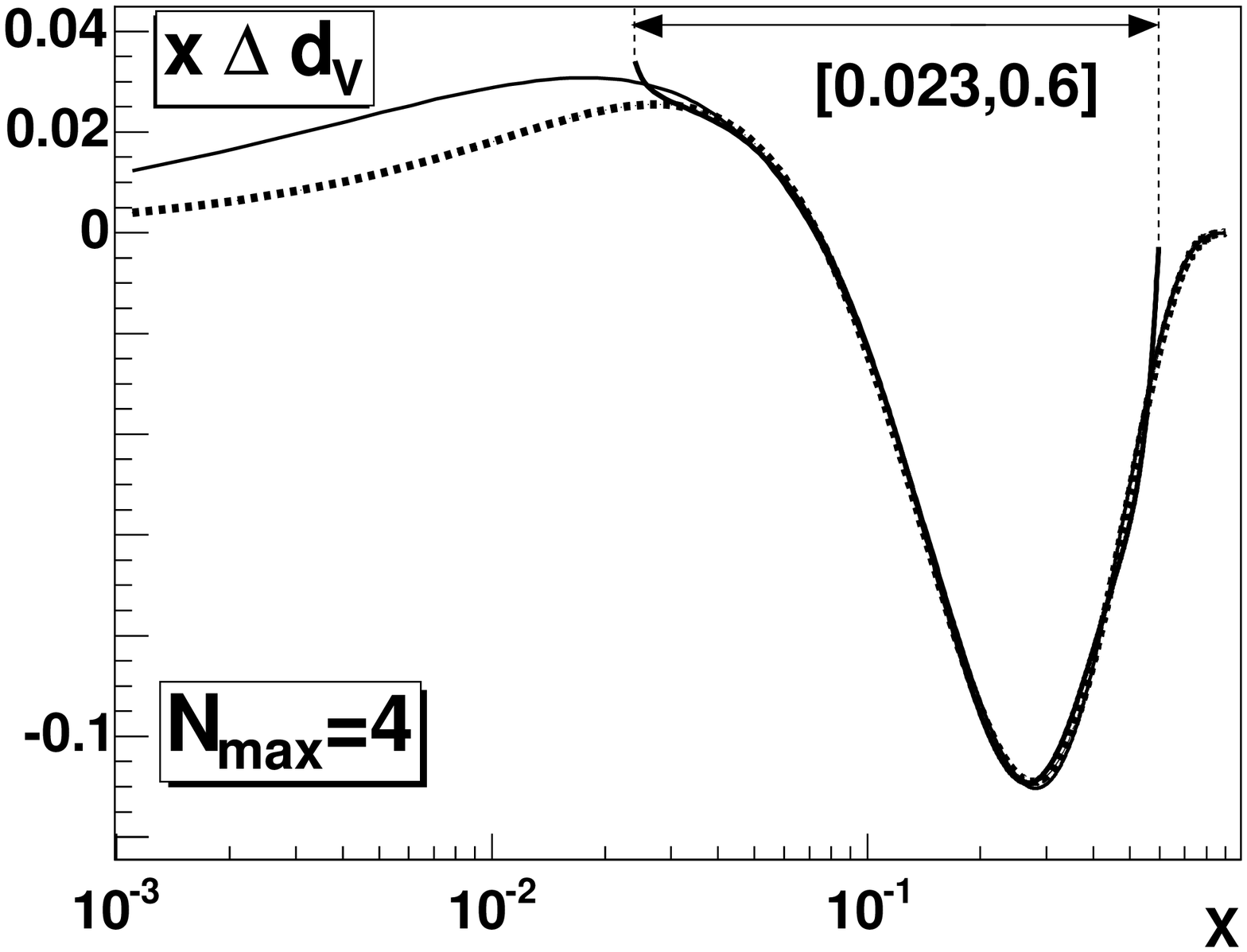}}
\end{center}
\end{figure}

\end{document}